\documentclass[aps,onecolumn,showpacs]{revtex4}
\usepackage{amsmath}
\usepackage{graphics}

\begin{document}

\title{Particular type of gap in the spectrum of multiband superconductors}

\author{P.I. Arseev $^{1,2}$  \thanks{e-mail: ars@lpi.ru}, S.O. Loiko$^{1}$ , N.K. Fedorov$^{1}$ }

\affiliation { $^{1}$ P.N.Lebedev Physical Institute of RAS, Moscow, Russia}
\affiliation { $^{2}$ National Research University, Higher School of Economics, Moscow, Russia }


\begin{abstract}
 We show, that in contrast to the free electron model (standard BCS model), a particular
gap in the spectrum of multiband superconductors opens at some distance from the Fermi energy, if conduction band is composed of hybridized atomic orbitals of different symmetries.
This gap has  composite superconducting-hybridization origin, because it exists only if both the superconductivity and the hybridization between different kinds of orbitals are present. So for many classes of superconductors with multiorbital structure such spectrum changes should take place.
These particular changes  in the spectrum at some distance from the Fermi level result in slow convergence of the spectral weight of the optical conductivity even in quite conventional superconductors with isotropic $s$-wave pairing mechanism.
\end{abstract}

\maketitle

The classical BCS model is based on the picture of free electron gas. This approach is adequate  for metals with wide conduction band  formed by identical atomic orbitals. In this case any changes of excitation spectra after transition to the superconducting state occur only within energy scale of a few $\Delta$ near the Fermi level. As a consequence, optical conductivity in a frequency range more than 4-6 $\Delta$ is practically the same in the normal and superconducting states. That was the case for the previously known superconductors, but for the new high-Tc families there are some evidences, that optical conductivity in superconducting state noticeably differs from normal conductivity to larger frequencies \cite{Basov,Molegraaf,SantanderSyro04,Boris04,Homes,Deutscher05,carbone,Charnukha_2011,Charnukha_2014}.

This fact can hardly be explained in the framework of ordinary "metallic" BCS model, but looking at the new superconductors from "semiconductor" point of view we can understand how changes of electron spectrum can occur rather far from the Fermi surface. "Semiconductor point of view" means that we take into consideration several bands near the Fermi level, paying attention to the symmetry of atomic orbitals. So several energy scales connected with the position of these bands and its hybridization value appear in the theory.
This is in contrast with the BCS theory, where the only parameter is the ratio $\Delta/E_F$.

Multiband models are now widely discussed for the new classes of superconductors beginning from YBCO, and more recent MgB2 and FeAs, FeSe families \cite{Mazin,FeAs,Hirschfeld,Dagotto,Chubukov1}.
Usually the effects of intra- and inter-band interactions attract main attention in multiband models ( e.g. \cite{Chubukov,Hotta,pairsym,SanoOno,Sadovskii_2008,Hirsch2016,2011Hirsch}). Various  signs and relative values  of these interactions lead to order parameters of different signs  at different pockets (sheets) of the Fermi surface.
In the present paper we are interested in quite different effect which takes place even if there is a single Fermi surface and for which interplay of different interactions between bands is of no importance.

To understand, what unusual for the BCS model spectrum changes are encountered in multiband systems,
we consider the two-band model of a superconductor described by the Hamiltonian

\begin{eqnarray}
\label{hamilt}
{H} & =  & {\sum_{{\mathbf{k}},\alpha
}\xi_{a}({\mathbf{k}})a_{{\mathbf{k}},\alpha
}^{+}a^{}_{{\mathbf{k}},\alpha }+\sum_{{\mathbf{k}},\alpha
}\xi_{c}({\mathbf{k}})c_{{\mathbf{k}},\alpha
}^{+}c^{}_{{\mathbf{k}},\alpha }}    +  \nonumber \\
& + &
{ \sum_{{\mathbf{k}},\alpha
}(t_{ac}({\mathbf{k}})a^{+}_{{\mathbf{k}},\alpha}
c_{{\mathbf{k}},\alpha}+h.c.)} -
                                                 \nonumber \\
      &  -  &{
\sum_{{\mathbf{k}}}\left(\Delta_{a}a_{-{\mathbf{k}}\downarrow}^{+}
 a_{{\mathbf{k}}\uparrow}^{+}+h.c.\right)} - \nonumber \\
  & - &  {\sum_{{\mathbf{k}}}\left(\Delta_{c}c_{-{\mathbf{k}}\downarrow}^{+}
 c_{{\mathbf{k}}\uparrow}^{+}+h.c.\right)}
\end{eqnarray}
where $a_{{\mathbf{k}},\alpha }^{+}$ and $c_{{\mathbf{k}},\alpha }^{+}$ are creation operators for electrons with the quasimomentum ${\mathbf{k}}$ and spin $\alpha$ in bands $a$ and $c$, respectively, which are generated by atomic orbitals with different kinds of symmetry;
$\xi_{a}({\mathbf{k}})$ and $\xi_{c}({\mathbf{k}})$ are the electron energies in the respective bands, and $t_{ac}({\mathbf{k}})$  is the matrix element corresponding to the single-particle interband hybridization \cite{our_cond}. The last two terms in eq.(\ref{hamilt}) describe the electron-electron interaction in the bands. They are written directly in the form corresponding to the mean-field approximation.
Of course in the multiband superconductors all types of pairing usually exist, but the interband pairing is not crucial for the effect we should like to discuss.  So to make the idea more clear we omit in Hamiltonian (\ref{hamilt}) interband pairing terms.

The anomalous averages $\Delta_a$ and $\Delta_c$
introduced here are defined in the usual way:
\begin{eqnarray}
\label{ordpar_init}
\Delta_a&=&-U_a\frac{1}{N}\sum_{{\mathbf{k}}}\langle
a_{{\mathbf{k}}\uparrow} a_{-{\mathbf{k}}\downarrow} \rangle
\nonumber \\
\Delta_c&=&-U_c\frac{1}{N}\sum_{{\mathbf{k}}}\langle
c_{{\mathbf{k}}\uparrow} c_{-{\mathbf{k}}\downarrow} \rangle
\end{eqnarray}

Since a specific mechanism responsible for the superconducting pairing between electrons is not crucial for our aims, we assume for simplicity that the interaction in the both bands is written as in Gor'kov's equations. As a result, the initial order parameters given by eqs. (\ref{ordpar_init}) in the $a$ and $c$ bands are isotropic; i.e., they do not depend on the momentum.
For superconductivity to appear it is sufficient to have attraction only in one band. The electron-electron interaction can even be repulsive in the other band as long as this band is much wider (with lesser density of sates). Different signs of the electron-electron interactions within the bands lead to the effective order parameters which are alternating-sign functions of the quasimomentum \cite{our_cond}. At present we need only, that interaction in one or two bands is sufficient for superconductivity to exist.

Because the main part of new superconductors are quasi 2D systems, in what follows we assume that Hamiltonian (\ref{hamilt}) describes 2D superconductor and present pictures of 2D excitation spectrum, though the gap of the new type exists in 3D system as well.

The excitation spectrum of the model in the superconducting state consists of two branches and has the form

\begin{align}
\label{spectrum}
 E_{\pm}^2({\bf k}) =\frac{\varepsilon_a^2({\bf k})+\varepsilon_c^2({\bf k})}{2} \pm
  \frac{1}{2}
\sqrt{(\varepsilon_a^2({\bf k})-\varepsilon_c^2({\bf k}))^2 +
4t_{ac}^2({\bf k})[(\xi_a({\bf
                        k})+\xi_c({\bf k}))^2 + (\Delta_a-\Delta_c)^2]}
\end{align}

where we denote
\begin{align}
\varepsilon_a^2({\bf k}) &= \xi_a^2({\bf k})+\Delta_a^2 +
t_{ac}^2({\bf k}), \nonumber \\
\varepsilon_c^2({\bf k})& = \xi_c^2({\bf k})+\Delta_c^2 + t_{ac}^2({\bf k}).\nonumber
\end{align}

As opposed to the standard BCS model with the single gap this spectrum has two gaps appeared due to the superconductivity. One opens as usual at the Fermi surface and the second one separates two branches of the spectrum $E_{+}({\bf k})$ and $E_{-}({\bf k})$ near the line $\xi_a+\xi_c=0$. The value of this gap can be estimated from eq.(\ref{spectrum})  using expansion in small parameter $|\Delta_c-\Delta_a|/\sqrt{t_{ac}^2({\bf k})+\xi_c^2({\bf k})}$
\begin{align}
\label{spectrum_m}
E_{\pm}({\bf k}) &\approx \sqrt{t_{ac}^2(\bf k)+\xi_c^2({\bf k}) \pm |t_{ac}({\bf k})||\Delta_c-\Delta_a|} \nonumber \\
& \approx \sqrt{t_{ac}({\bf k})^2+\xi_c^2({\bf k})} \pm \frac{|t_{ac}({\bf k})||\Delta_c-\Delta_a|}{2\sqrt{t_{ac}^2({\bf k})+\xi_c^2({\bf k})}}
\end{align}
We see, that this gap appears  due to both superconductivity and hybridization and vanishes if either $t_{ac}$ or $\Delta$ equals to zero. We can call it a "composite" gap, because hybridization level repulsion and superconductivity make equal contribution to its formation. To avoid misunderstanding we should stress that this gap is not a total energy gap in the density of states, contrary to the gap at the Fermi surface. It separates  two branches of the excitation spectrum and its value and position changes along the line
$\xi_a(\bf k)+\xi_c(\bf k)=0$ in the Brillouin zone.

\begin{figure}[h!]
\resizebox{0.75\textwidth}{!}{%
 \includegraphics{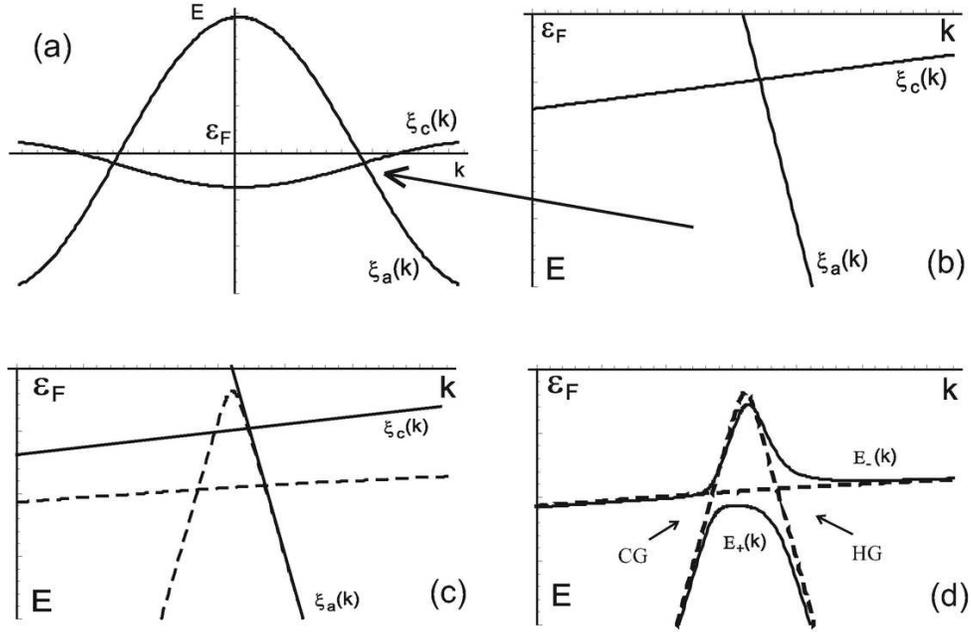}
 }
\caption{Schematic view of the initial two band picture with superconductivity switched off and without interband hybridization (a) and zoom of the area of the initial bands intersection (b).
The area of the intersections of energy spectrum branches:  (c) with superconductivity switched on  and without hybridization (dashed lines); (d) with both superconductivity and hybridization switched on (solid lines).
The ordinary hybridization gap (HG) and new composite gap (CG) are indicated by the arrows.
}
\label{comp_gap}
\end{figure}

Simple 1D section of the spectrum shown in fig.\ref{comp_gap} helps us to understand in what part of the spectrum and due to what reason this new gap opens. In the normal state there is only one crossing of the two bands (fig.\ref{comp_gap}b). Degeneracy in this point is removed in a usual way by the hybridization $t_{ac}$ (not shown in the figure \ref{comp_gap}b).  But after superconducting transition the quasiparticle "mirror" branch of the wide band crosses the quasiparticle branch of the narrow band once more as it is depicted in fig.\ref{comp_gap}c.
Now the hybridization $t_{ac}$ splits the spectrum also at this second intersection point besides the ordinary avoided crossing band splitting (fig.\ref{comp_gap}d). That is why this new gap has some composite character: without superconductivity there is no second intersection point  and without hybridization  no gap opens at the band crossing points.

This simple qualitative explanation is quite correct in the case if the order parameter is nonzero only in one of the bands. If the order parameters are nonzero in the both bands the situation is qualitatively the same but a little bit more complicated in details, because superconducting interaction connects branches under and above the Fermi level, so near the intersection we have a system of four but not of two equations (corresponding to four connected quasiparticle branches). This specific character of superconductivity equations leads to the result that in a completely symmetric situation for $\Delta_a=\Delta_c$ (and $\xi_a=-\xi_c$) this new gap vanishes, as it can be seen in eq. \ref{spectrum_m}.

Figure \ref{comp_gap}d shows the case of very small hybridization value ($t\approx \Delta$) in order to illustrate the effect clearly. If hybridization increases the quasiparticle spectrum near the line, where the bands intersect in the normal state , deforms significantly and only the new gap, which is much smaller, can be distinctly identified in the spectrum.

\begin{figure}[h!]
\resizebox{0.75\textwidth}{!}{
\includegraphics{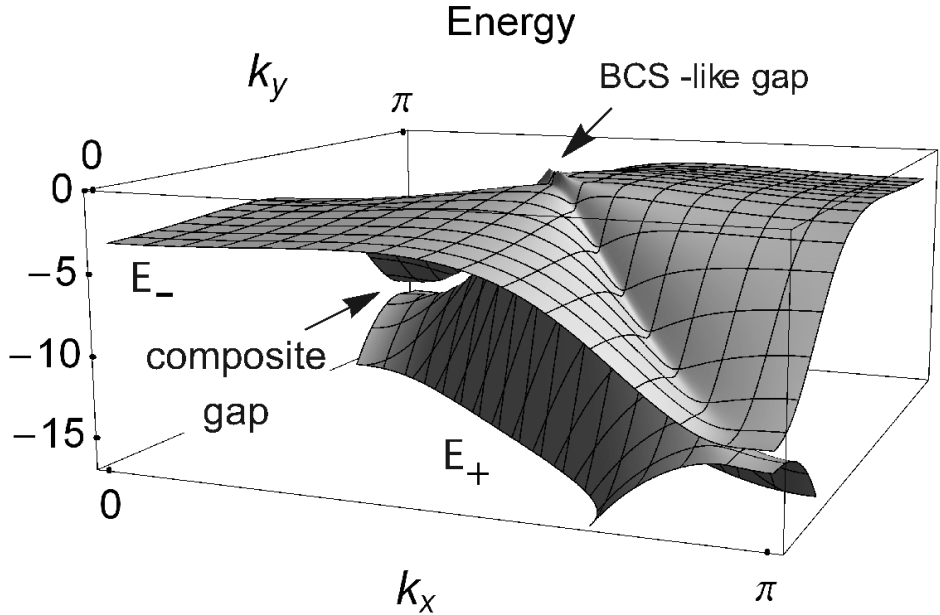}
}
\caption{$E_{-}(\mathbf{k})$ and $E_{+}(\mathbf{k})$ branches of the excitation spectrum lying under the Fermi level in superconducting case. A quarter of the Brillouin zone is shown. }
\label{spectr}
\end{figure}

For some more real choice of the parameters the total spectrum is shown in fig. \ref{spectr}.
The calculations were performed for the two bands with the dispersion laws typical for the simple cubic lattice, namely  $\xi_a(\mathbf{k})=t_a (\cos k_x+\cos k_y)$ and $\xi_c(\mathbf{k})=\varepsilon_{c0}+t_c (\cos k_x+\cos k_y)$.
Since it is assumed that these bands are formed by the atomic orbitals of different types of symmetry, the matrix element of hybridization between them is a strongly anisotropic function of quasimomentum  $t_{ac}(\mathbf{k})=t_{ac}^0(\cos k_x-\cos k_y)$ \cite{our_cond}. (It is not crucial for the discussed effect but it is closely related to our previous results \cite{our_JETP_Lett}).
The calculations were performed with the following values of the parameters of the two-band model: $t_a=30$, $t_c=-1$, $t_{ac}^0=5$, $\Delta_a=-0.18$, $\Delta_c=0.81$ (in units of $|t_c|$). For such values of parameters $\Delta$ is much less than the bandwidth and hybridization matrix element is less than the bandwidth but large enough compared to the order parameter.

\begin{figure}[h!]
\resizebox{0.75\textwidth}{!}{
\includegraphics{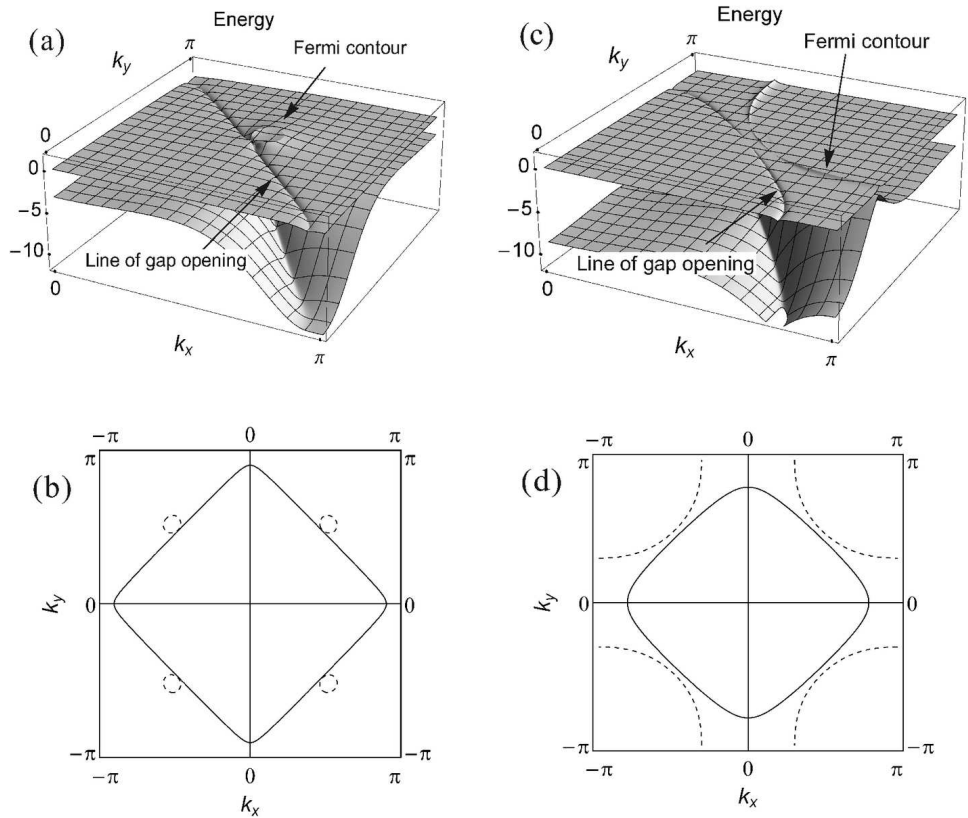}
}
\caption{$E_{-}$ branch and the difference between quasiparticle
energies $E_{-}$ for the superconducting and normal states as
functions of $\mathbf{k}$ in a quarter of the Brillouin zone (a), and the Brillouin zone with the
lines along which the difference has its maximum value, i.e. the
Fermi contour (dashed line) and the $\xi_a+\xi_c=0$ contour
(solid line) (b) for $\varepsilon_{c0}=-0.5$; the same for $\varepsilon_{c0}=-5$ (c) and (d) correspondingly.  } \label{differ05}
\end{figure}

In order to show the area in the Brillouin zone where the spectrum changes substantially after superconducting transition we plot in figs. \ref{differ05} the difference  $\Delta E=E_-^s-E_-^n$ between the quasiparticle energies $E_-(k)$ in the superconducting and normal states.
For the usual BCS model such plot gives the ridge of the $\Delta$ height at the Fermi surface which quickly tends to zero: $\Delta E(k)=\sqrt{\xi^2(k)+\Delta^2}- \xi(k)\simeq (\Delta^2/\xi^2(k))$ as $\xi(k)>\Delta$.
The distinctive feature of the spectrum of the two-band model is significant change in the quasiparticle energy not only near the Fermi contour (displayed in figs.\ref{differ05}b, \ref{differ05}d by dashed line), but also along the line $\xi_a+\xi_c=0$ (displayed in figs.\ref{differ05}b, \ref{differ05}d by solid line).

In the first case the narrow band is near the center of the wide band: $\varepsilon_{c0}=-0.5$ (in units of $|t_c|$) and in the second case the narrow band is moved further: $\varepsilon_{c0}=-5$. In the second case the intersection line of the two bands is moved further from the Fermi level, which is located in our calculations in the middle of the wide band. The Fermi surfaces in these two cases are quite different, nevertheless  in any case we see two lines of substantial quasiparticle energy changes instead of one line for the BCS model.

The impact of this particular gap on physical observables depends on the shape of the spectrum and on the position of this gap.
Especially important is how close the new gap lies to flat enough parts of the spectrum which produce Van Hove singularities.

\begin{figure}[h!]
\resizebox{0.75\textwidth}{!}{
\includegraphics{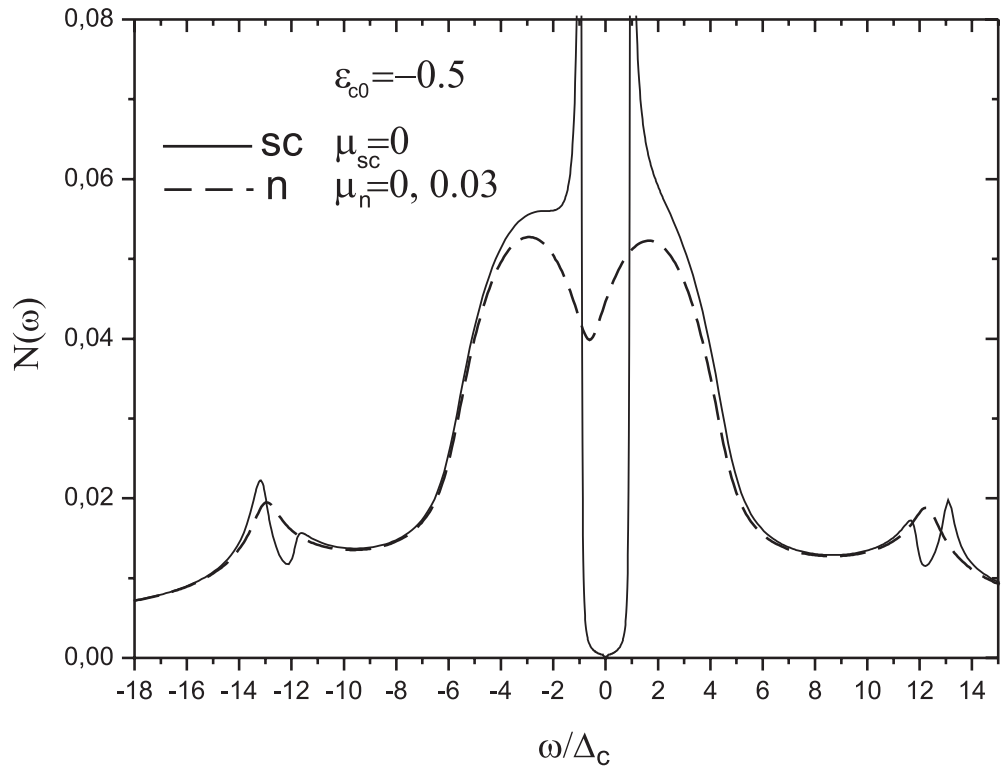}
}
\caption{Densities of states in the normal and superconducting states for the positions of the center of the narrow band with respect to the chemical potential $\varepsilon_{c0}=-0.5$. Dashed line corresponds to the normal state and solid line --- to the superconducting state. }
\label{dens}
\end{figure}

In fig.\ref{dens} the density of states is shown for the two band model in the normal and superconducting states for the case $\varepsilon_{c0}=-0.5$. For the given above choice of parameters  this new gap opens near Van Hove singularities at the edge of the Brillouin zone. In this case the effect is very prominent. For the superconducting state we see the usual BCS like gap structure at the Fermi level and then the density of states quickly approaches the normal density of states. But suddenly at the energies of $(12\div 14) \Delta$ from the Fermi level there are again strong deviations between the superconducting and normal densities of states. It is the effect of the new composite gap opening.

\subsection*{Conclusions}

So we have demonstrated that in multiband superconductors there exists a simple mechanism which leads to opening a gap of a new "composite" type in the spectrum. This spectrum changes can take place  rather far from the Fermi surface.
Qualitatively the mechanism is connected with mirror reflection of quasiparticle branches after the superconducting transition and repulsion the spectrum branches along new intersection lines due to the hybridization.

This point (line,surface) where quasipartical branch of one band, remaining in the normal state, crosses superconductor "mirror" quasipartical branch of another band, vanishing in the normal state, determines the position of the new gap line.
The value of the gap is given by equations like eq.\ref{spectrum_m} and is proportional both to the hybridization energy $t_{ac}$ and to the order parameter(s)   $\Delta$.

Of course if intersection point moves far from the Fermi level the value of this gap decreases. Nevertheless eq.\ref{spectrum_m} shows that it diminishes as  $t_{ac}\Delta/\xi_{c} $ where $\xi_{c}$ approximately determines the distance of the gap position from the Fermi level.
It can be easily estimated that relative changes of the density of states and conductivity are also of the order of this parameter $t_{ac}\Delta/\xi_{c}\simeq t_{ac}\Delta/(E-E_F)$.
This linear dependence on the parameter $\Delta/(E-E_F)$ is much slower than usual for the BCS model square-low dependence $(\Delta/(E-E_F))^2$.

Though this particular gap in the spectrum does not result in a total gap in the density of states it is very important for any processes of absorbtion or emission of light by the multiband superconductors. We see in Figs. \ref{spectr},  \ref{dens} that conditions of absorption or emission change after superconducting transition
not only for photons with frequency of the order of $\Delta$ but also for photons with frequency $(12\div 14) \Delta$ .
This means, for example, that optical conductivity should change noticeably in a spectral range much greater than $\Delta$  in quite conventional superconductors with initial isotropic $s$-wave pairing mechanism which is in accord with mentioned above experimental studies \cite{Basov,Molegraaf,SantanderSyro04,Boris04,Homes,Deutscher05,carbone,Charnukha_2011,Charnukha_2014}.
It is just this general mechanism of the composite gap opening, that qualitatively explains our previous results for slow increase of the conductivity spectral weight with frequency cutoff growing in the two-band model \cite{our_JETP_Lett}.

\end{document}